# Locally Enhanced Image Quality with Tunable Hybrid Metasurfaces

Alena V. Shchelokova[1], Alexey P. Slobozhanyuk[1,2,*], Irina V. Melchakova[1], Stanislav B. Glybovski[1], Andrew G. Webb[3], Yuri S. Kivshar[1,2], and Pavel A. Belov[1]

[1] Department of Nanophotonics and Metamaterials, ITMO University, St. Petersburg 197101, Russia

[2] Nonlinear Physics Center, Australian National University, Canberra ACT 2601, Australia

[3] Department of Radiology, Leiden University Medical Center, Leiden 2333 ZA, Netherlands

[*] Correspondence and requests for materials should be addressed to A.P.S. (email: aleksei.slobozhaniuk@anu.edu.au)

## Abstract

Metasurfaces represent a new paradigm in artificial subwavelength structures due to their potential to overcome many challenges typically associated with bulk metamaterials. The ability making very thin structures and change their properties dynamically make metasurfaces an exceptional meta-optics platform for engineering advanced electromagnetic and photonic metadevices. Here, we suggest and demonstrate experimentally a novel tunable metasurface capable to enhance significantly the local image quality in *magnetic resonance imaging* (MRI). We present a design of the hybrid metasurface based on electromagnetically-coupled dielectric and metallic elements. We demonstrate how to tailor the spectral characteristics of the metasurface eigenmodes by changing dynamically the effective permittivity of the structure. By maximizing a coupling between metasurface eigenmodes and transmitted and received fields in the MRI system, we enhance the device sensitivity that results in a substantial improvement of the image quality.

## I. INTRODUCTION

Metasurfaces [1, 2, 3] are artificial subwavelength-patterned structures representing an ultrathin planar realisation of the metamaterial concept [4]. An essential advantage of metasurfaces over metamaterials is their simplicity of manufacturing and ability to control reflected and transmitted waves, as well as to specifically sculpt the near field patterns over a frequency range from radio to optical [5, 6, 7]. Structural elements of metasurfaces can be made from different metallic [8], plasmonic [9] or low loss dielectric materials [10], while their geometrical dimensions and elements separations are essentially subwavelength. In this way, homogeneous or nearly homogeneous metasurfaces reflect and transmit plane waves as infinitely thin sheets, supporting both electric and magnetic currents [5]. Metasurfaces have been employed for different applications including flat optical components [11, 12, 13, 14], holograms [15, 16], vortex generators [17], optical information processing [18], nonreciprocal components [19], chirality enhancement [20], demonstration of optical spin-orbit interactions [21, 22, 23], and control of surface waves [24].

A majority of these metasurface structures has fixed properties, e.g. operational bandwidth or functionality: but active control of metasurfaces is highly desirable [25]. Recently, several groups reported different methods for a design of metasurfaces with tunable responses [6]. Metasurfaces can also display extreme nonlinear responses [26, 27] that enable full optical control and tunability of material parameters as well as being able to achieve efficient generation of second [28] and third harmonics [29]. As a result, these two significant properties, ultrathin planar geometry and dynamic response made metasurfaces an exceptional platform for engineering novel photonic metadevices [25].

One important application of metasurfaces is the image quality enhancement that is crucially important in different areas including biology, medicine, and nanotechnologies. In particular, it has been demonstrated experimentally that a metasurface lens designed for optical frequencies can be used to provide diffraction-limited focal spots [14] and to resolve features with subwavelength spacing, while parity-time symmetric nonlocal metasurfaces allow ideal aberration-free optical images [30]. Another imaging area where metasurfaces can potentially be extremely useful is magnetic resonance imaging (MRI), one of the most important clinical modalities for detection of various diseases in humans [31]. MRI image quality and spatial resolution are ultimately limited by the signal to noise ratio (SNR) obtainable in a relevant clinical scanning timeframe [32]. Recently a new conceptual idea has been suggested to substantially increase SNR of commercial MRI with the aid of microwave metasurfaces made by resonant metallic inclusions [33]. This effect has been further advanced in in-vivo imaging and spectroscopy of the human brain [34]. However, all previous designs produced a locally enhanced magnetic field that was distinctly non-uniform over the extension of metasurface. Also, fine tuning of the spectral response was impractical. Therefore, compact tunable designs are still very appealing for practical applications.

Here, we propose and realise experimentally a novel type of metasurfaces capable to enhance significantly the image quality applicable to MRI. To overcome the problem of relatively large physical dimensions, inspired by the recent advances in the development of bilayer metasurfaces [35] for visible light, we employ a hybrid microwave metasurface design based on electromagnetically coupled components of both dielectric materials and metallic elements. In addition, by changing the effective permittivity of the structure we observe a substantial spectral shift of the metasurface eigenmode and can fine-tune its properties for patient-specific MRI experiments. In contrast to previous designs [33, 34], where the specific near-field pattern of the metasurface eigenmode has a sinusoidal shape of the radiofrequency (RF) magnetic field along its length, and therefore results in an inhomogeneous SNR enhancement and non-ideal images, here by altering the coupling strength between the dielectric and metallic layers we are able to achieve different shapes of the near field mode profile, including an almost homogeneous one in the region-of-interest.

## II. DESIGN OF METASURFACES

In order to realize a novel tunable hybrid metasurface, we start from the microwave metasurface based on resonant elongated metallic particles inserted into a dielectric matrix with a permittivity $\varepsilon$ (Fig. 1a). Such a metasurface supports a number of eigenstates, which arise due to the splitting of the original resonant frequency into several bands, induced by giant coupling between many resonant particles placed at subwavelength distances [36, 37, 38, 39, 40]. In particular, a very interesting phenomenon is observed in the range of frequencies within the first Fabry-Pérot resonance where it is possible to enhance and spatially redistribute the magnetic and electric fields since they correspond to the near-field region, where the electric and magnetic fields distribution of the same eigenstate are decoupled [36]. This property can be employed to increase SNR in MRI [33]. The sensitivity of the MRI receiver is directly proportional to the ratio of the RF magnetic field in the scanned region to the square root of the absorbed power, and the metasurface increases this sensitivity.

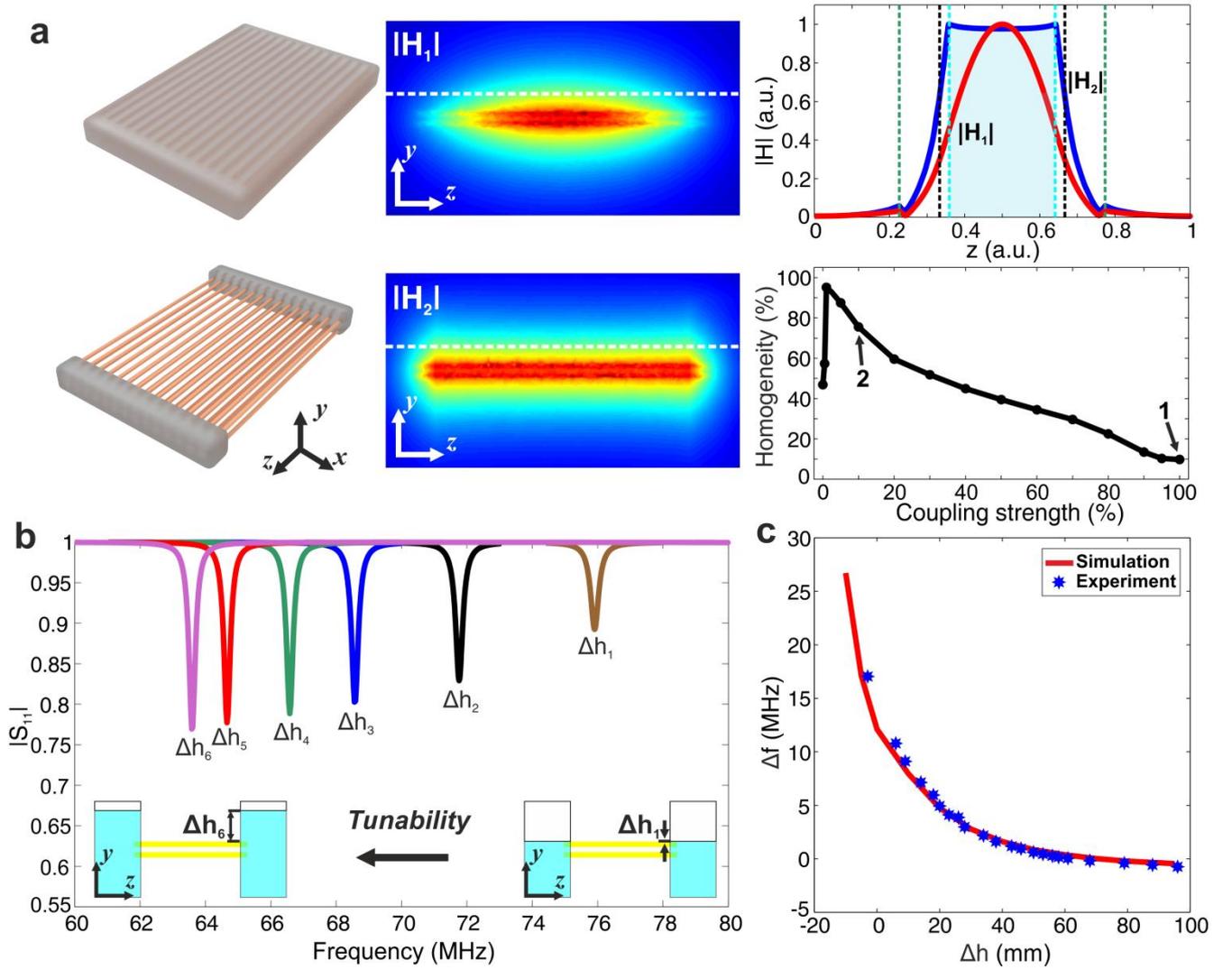

FIG. 1. Basic concept and electromagnetic properties of tunable hybrid metasurfaces. (a) A comparison between the metasurface from the previous study [33] realized by an array of 14x2 brass wires placed inside a homogeneous dielectric medium (top panel) and the new hybrid metasurface (bottom panel). The central panel demonstrates 2D maps of the magnetic field distribution in the *yz* plane for both cases. The right top panel demonstrates the near field mode profile of the first eigenmode of different metasurfaces calculated along white dashed lines (shown in the central panel) at the distance of 15 mm above the wires. The magnetic near-field profile of the first eigenmode of the hybrid metasurface has a flat shape in comparison with the field pattern observed in another type of metasurface which has an inhomogeneous sinusoidal form within the region of interest (shown by shaded area and light blue dashed lines). The boundaries of the wires are shown by black dashed lines, the boundaries of dielectric blocks with green lines. The right bottom panel shows the dependences of the RF magnetic field homogeneity in the region-of-interest as a function of coupling strength. The two markers indicate field homogeneity in homogeneous (1) and hybrid (2) designs. The homogeneity level has been determined relative to the metasurface length and is represented as the region of space where RF magnetic field amplitude variations are less than 1%. (b) Numerical simulation results of the reflection coefficient ($|S_{11}|$) of a small loop antenna (with a radius of 40 mm) placed above the hybrid resonator at a 100 mm distance from the wires as a function of the dielectric level ($\Delta h$): $\Delta h_1$=0 mm; $\Delta h_2$=10 mm; $\Delta h_3$=20 mm; $\Delta h_4$=30 mm; $\Delta h_5$=50 mm; $\Delta h_6$=80 mm. The insets schematically illustrate the hybrid metasurface geometry and the tunability principle. c) Resonance frequency ($\Delta f=f_{resonance}-$63.8MHz) as a function of the dielectric level ($\Delta h$).

In the simulations a linearly polarized plane wave excites a set of transverse eigenmodes near the first Fabry-Pérot resonance and the lowest mode has a maximum of the magnetic field in the central region of the metasurface (Fig. 1a), with the electric field maximum being located near the edges of the wires. The RF magnetic field pattern along the metasurface has a sinusoidal-like shape (right top panel on the Fig. 1a) analogous to the conventional standing waves in optical cavities [41]. This is problematic for MR images since the signal intensities and image contrast are no longer homogeneous across the region-of-interest, a phenomenon which was observed with previous designs [33, 34]. However, this problem can be solved by realizing a hybrid metasurface structure where only the edges of the wires are loaded with high permittivity dielectric material. Due to the very high contrast in permittivity between two regions, the middle part of the metasurface embedded in low epsilon material $\varepsilon_1$ while the edges are in a high permittivity material with $\varepsilon_2$, the magnetic field can be extended and its shape becomes almost homogenous (cf. top and bottom central panels on the Figure 1a). At the same time the electric field is still mostly localized inside the high permittivity blocks. Thus, it is possible *to achieve a homogeneous local RF magnetic field enhancement*, while *minimizing* the electric field amplitude (and therefore the specific absorption rate) in the region of interest. Furthermore, by changing the coupling strength via varying the length of wires inserted in the dielectric blocks (right bottom panel on the Fig. 1a), almost any slope of the RF magnetic field pattern can be designed, including realisations of different spatial domains with magnetic field homogeneity around 98%. It is also very important to mention that due to a hybrid nature of the metasurface its dimensions are much smaller than the operational wavelength of the MRI machines (see detailed discussions in Supplemental Material [42])

Different modes of the hybrid metasurface have different penetration depths [33] and for the current study the lowest transverse eigenmode, which has the deepest penetration depth and the most homogeneous in-plane magnetic field pattern, has been employed. In order to achieve tunability of the metasurface we externally change the effective permittivity of the dielectric blocks. For the proof-of-principle experiment we used distilled water as a high permittivity material and realized the metasurface coupled to two dielectric blocks from both edges (see the inset on the Figure 1b and Appendix A for details). To demonstrate the effect of eigenmode tunability we have numerically calculated the reflection coefficient of a small loop antenna placed above the hybrid metasurface (Fig. 1b). For relatively small values of metasurface effective permittivity, the eigenmode is tuned to a higher frequency (light blue curve in the Figure 1b). By increasing the effective permittivity (e.g. by changing the water level ($\Delta h$)) the resonance frequency of the first eigenmode shifts to the lower frequency range, the higher the permittivity - the larger the shift, and in this way the device can be effectively tuned to 63.8 MHz, the working frequency of a 1.5 T MRI system (see Appendix B for further details on the numerical simulations). The effect can be explained by the fact that the eigenmode of the metasurface experiences an effective permittivity $\varepsilon_{eff}$ which depends on the volume fraction of the dielectric inclusions ($\varepsilon_1$ and $\varepsilon_2$) in the mixture [43] in the region close to the resonator and by changing this fraction to a certain extent it is possible to tune the resonance. The resonance shift saturates for some values of $\varepsilon_{eff}$ (water level $\Delta h$) due to the finite penetration depth of the eigenmode (Fig. 1c).

To confirm these dependences experimentally we have realized a hybrid metasurface as an array of nonmagnetic metallic wires coupled to the dielectric blocks, fillable by water (schematically shown in the inset of Figure 1b, see Appendix A for details). We gradually added water inside two hollow blocks and measured the reflection coefficient of a loop antenna placed above the metasurface as a function of $\Delta h$. Figure 1c demonstrates the ability to tune the metasurface resonance frequency. Excellent agreement is seen between the experimental measurements and the simulation results. It is worth noting that the MRI machine works in a narrow frequency band around the Larmor frequency (here for the Hydrogen imaging it is equal to 63.8 MHz) and the metasurface should be tuned exactly to this frequency. However, variation of electrical properties of a subject (a patient) located closely to the metasurface may shift the resonance from the Larmor frequency, decreasing the metasurface efficiency. Therefore, *the ability to easily fine-tune the resonant frequency is very useful from a practical standpoint, and a unique feature of this new metasurface design*.

## III. IMPACT OF METASURFACES ON RECEIVER SENSITIVITY

We have investigated the performance of the tunable hybrid metasurface on a 1.5 T MRI system. As a test object, we chose a rectangular phantom placed on the top of the metasurface. We used the birdcage body coil of the system, which is embedded in the bore of the scanner, for transmitting and receiving the RF signal. Figure 2 shows the ratio between SNR with the hybrid metasurface ($SNR_2$) and without it ($SNR_1$) as a function of $\Delta h$ calculated in the phantom. Initially, when the water level is below the level of the wires ($\Delta h<0$) no SNR enhancement is observed since the eigenmode is detuned. This also confirms that the metasurface itself does not increase the noise in the images. Then, by gradually increasing the water level, we observed a smooth increase in the SNR. The maximum SNR enhancement of 7.34 is achieved for $\Delta h=65$ mm which corresponds to the resonant frequency of the hybrid metasurface tuned to the Larmor frequency. The slight difference in the $\Delta h$ value in comparison with numerical simulation is due to imperfections in the shape of dielectric blocks. MRI scans of the phantom in an axial plane for two cases with a hybrid metasurface, tuned to the maximum efficiency, and without it are shown in the inset in Figure 2a. A large increase in the signal intensity can be clearly observed near the metasurface resonator. By further increasing $\Delta h$ the impact of the metasurface on the MR sensitivity becomes less pronounced due to the fact that coupling of the metasurface eigenmode to the Larmor frequency decreases. A very important advantage of the hybrid metasurface is an almost homogeneous in-plane profile of the RF magnetic field near the meta-surface that results in nearly uniform signal from the region-of-interest (Fig. 2b). Furthermore, the lowest eigenmode of the metasurface enables a high penetration depth of the enhanced transmit field, which leads to SNR enhancement up to a 150 mm distance from the resonator (Fig. 2c). Near the metasurface the gain is a factor of 8, while at the distance of 140 mm from the wires inside the phantom it is about 1.5. In order to investigate the increase in transmit efficiency, as well as to compare SNR with and without the metasurface using optimum RF transmit conditions for both, we acquired gradient echo images with different tip angles. The dependence of the received MR signal on the tip angle for a gradient-echo sequence is given by [32]:

$$Signal(r) \sim A(r) \cdot \frac{\sin(\lambda(r) \cdot \theta_{nominal})}{(1 - e^{-\frac{TR}{T_1(r)}}) \cdot \cos(\lambda(r) \cdot \theta_{nominal})} \quad (1)$$

where $A(r)$ is a spatial coefficient which includes local spin density, $B_1^-(r)$ is the receive sensitivity, $T_2$ the transverse relaxation time, $T_1(r)$ the longitudinal relaxation time, TR the repetition time and $\theta_{nominal}$ the tip angle which was set in the scan parameters. The transmit efficiency $\lambda(r)$ is directly proportional to the transmit RF magnetic field $B_1^+(r)$. When we considered only the phantom without a hybrid metasurface and used a birdcage coil for transmitting and receiving the MR signal we obtained an optimal tip angle (corresponding to an optimal applied RF power level) where there is a maximum in the received signal (blue curve in Figure 3). Generally, for a spin-echo sequence, the optimal tip angle is equal to $90^0$ and for a gradient-echo sequence, it is much lower than $90^0$ unless the TR is set to be at least three times the $T_1$ value, which would result in a very long scan time [32]. The *detuned metasurface has no effect on this dependence*, i.e. the signal dependence on the tip angle in this case coincides with the curve for the case without the metasurface. However, when the metasurface is tuned to a frequency close to 63.8 MHz (Fig. 3) the curves shift horizontally to the left, meaning that $\lambda(r)$ increases, or in other words, the metasurface enhances the transmitted RF magnetic field $B_1^+(r)$ in the region-of-interest for the same applied power.

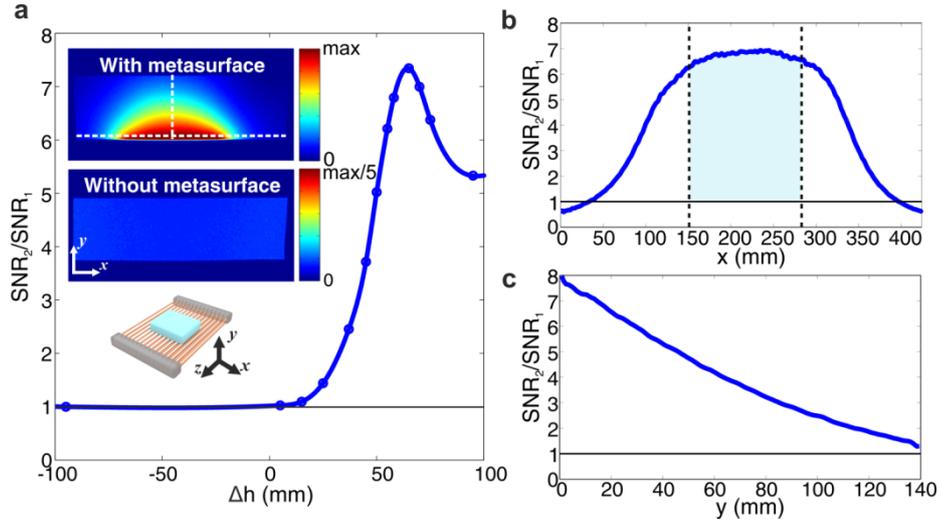

FIG. 2. MRI sensitivity vs. metasurface adjustment. a) Experimentally measured ratio between SNR with the metasurface (SNR$_2$) and without it (SNR$_1$) as a function of $\Delta h$. The insets correspond to the MR images of the phantom for the cases with the tuned metasurface present and without it and also shows the schematic position of the phantom on the metasurface. b,c) Experimentally measured ratio between SNR with metasurface and without it along the *x* (b) and *y* (c) directions in the phantom (shown by white dashed lines in the inset of panel (a)). The black dashed lines correspond to the wire boundaries and the blue shaded area to the region-of-interest.

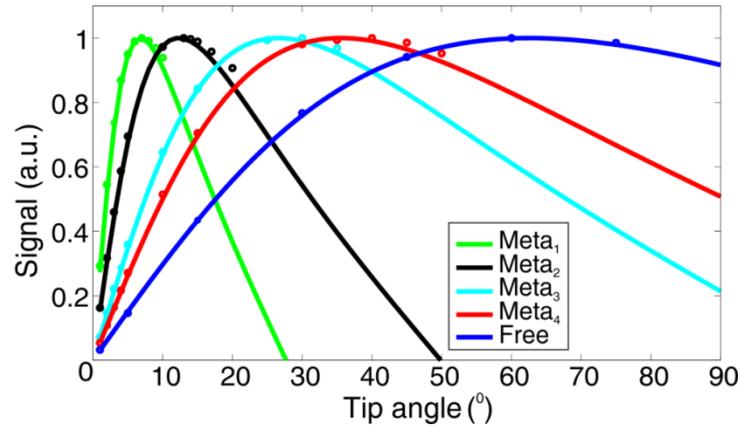

FIG. 3. MR signal vs. tip angle (applied RF power) and coupling strength between an artificial resonator and nuclear magnetic resonance frequency. $Meta_1$ was tuned to 64MHz, $Meta_2$ – 66.3MHz, $Meta_3$ – 71.2MHz, $Meta_4$ – 74.7MHz. The circles correspond to the experimental data and solid lines represent the theoretical results. Amplitudes were normalized by the following coefficients: $A_1$=2175 for $Meta_1$; $A_2$=349 for $Meta_2$; $A_3$=133 for $Meta_3$; $A_4$=96 for $Meta_4$; $A_5$=53 for a phantom without the metasurface.

An ability to locally enhance transmitted RF magnetic field in the region of interest leads to the global reduction of applied RF power and improves the safety of the patient. Additionally, due to the reciprocity principle, the metasurface increases the receive magnetic field $B_1^-(r)$, and the amplitude of the curves is higher (note that the values shown on Figure 3 are normalized). This directly confirms the

SNR enhancement due to an increase in the receiving sensitivity via the coupling with the metasurface eigenmode. The maximum effect is observed when a hybrid metasurface is tuned to 63.8 MHz.

## IV. ENHANCED IMAGE QUALITY

MRI quality depends on both the spatial resolution and SNR. The MRI system can reach subwavelength spatial resolution since typically the voxel sizes are far smaller than the wavelength, e.g. details with the sizes around 0.5 mm can be resolved at 1.5 T whereas the wavelength at 63.8 MHz is 4.7 m. This is achieved with the aid of gradient coils which produce a spatially variable magnetic field in addition to the main static magnetic field: this results in very slight modifications in the spatial distribution of the Larmor frequencies over the volume and results in spatial encoding of the received signal [32]. Higher spatial resolution results in the reduction of SNR due to the lower number of protons per voxel [32], and this reduction in SNR ultimately limits the spatial resolution of the image. It is important to note that small features can be visualized by time averaging the signal through multiple scans, but this results in very long scanning times which are not practical in human applications.

The metasurface improves SNR, so the image quality should also be enhanced. We perform several MRI scanning sessions by using a grapefruit (Fig. 4) and a birdcage coil for transmitting and receiving the RF signal. In the absence of the metasurface, it is possible to achieve an MR image of a grapefruit with a low spatial resolution using a large field-of-view (FoV) (see Fig. 4, top left and middle images), while for smaller voxels (smaller FoV) SNR becomes very low, so that image details such as segments of the grapefruit cannot be resolved properly (see Fig. 4, top right image). By placing the hybrid metasurface under the grapefruit, we obtain higher SNR resulting in an enhanced image quality allowing higher resolution scans to be acquired with adequate SNR to see more features. In this particular case, more informative images of a pulp structure can be acquired, together with a clear separation of the grapefruit segments, as well as details of a boundary between the pulp and peel (cf. right top and middle images in Fig. 4).

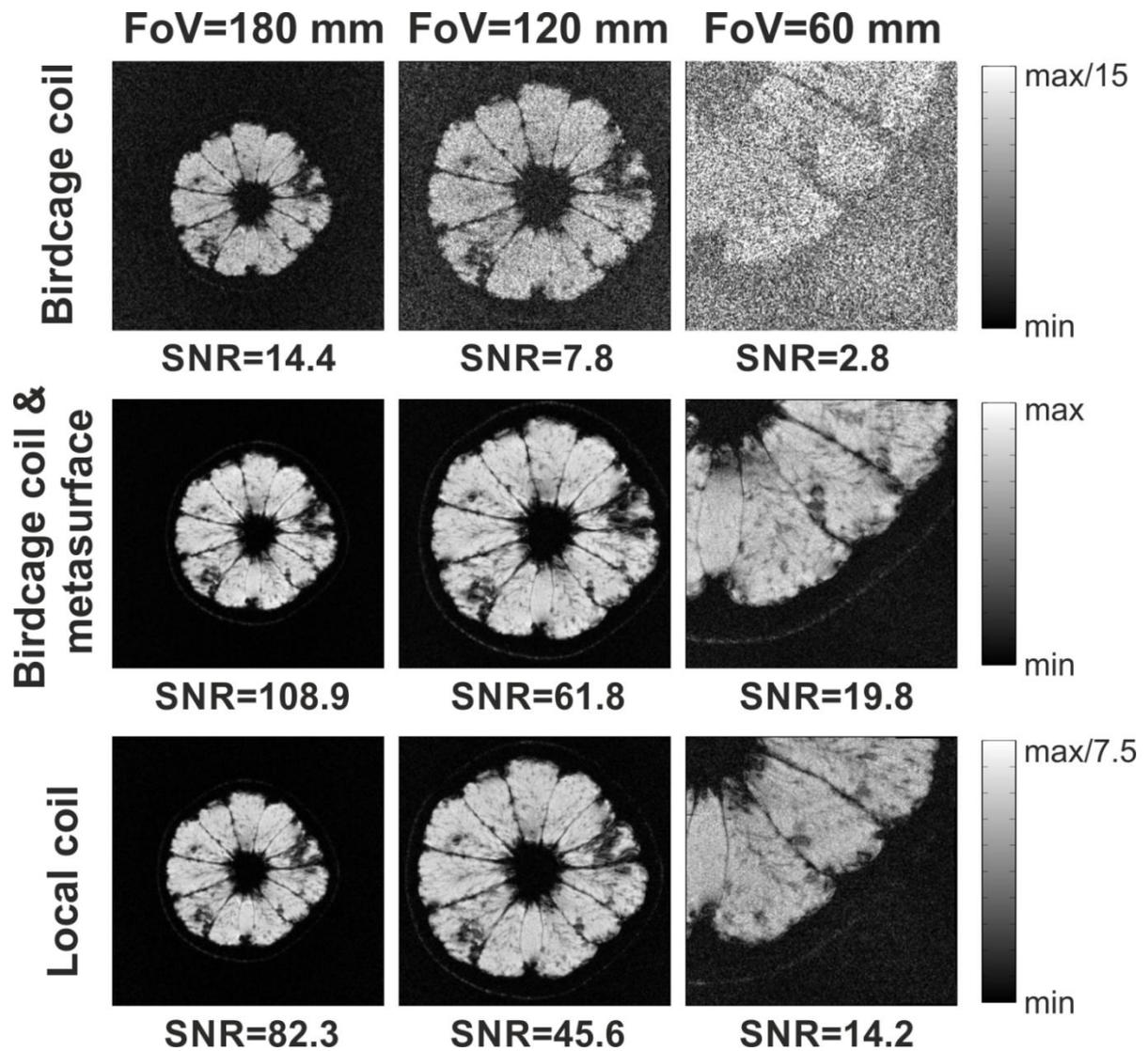

FIG. 4. Experimental demonstration of improved image quality with hybrid metasurfaces. MR images of grapefruit obtained with the birdcage coil only (top panel), with the birdcage coil and metasurface (middle panel) and with a 4-channel local receive coil (bottom panel) for different field-of-views (FoVs) corresponding to different voxels dimensions.

However, in clinical MRI, a signal from a patient is received by an array of closely fitted local receiver coils rather than much larger single birdcage body coil, which is typically used for transmit. Since the birdcage coil is far from the body, the filling factor is poor and so is the receive sensitivity. Therefore, we have compared a hybrid metasurface with a commercial dedicated 4-channel flexible receive coil

(Figure 4 bottom panel). The quality of image obtained with the metasurface is comparable with the image received by state-of-the-art local coil (cf. middle and bottom panels in Figure 4). However, the hybrid metasurface enables us to increase the SNR. Figure 4 (left top panel) shows the MRI image of a grapefruit with a transmit/receive birdcage coil: the SNR in the region of interest is 14.4. Next, the local receive coil has been placed under the object, and an SNR enhancement of 5.7 times in comparison with a birdcage coil was obtained, due to the higher sensitivity of the receive coil and its placement in the close vicinity of the object. Finally we used a hybrid metasurface and achieved SNR increase of 7.6 times versus the birdcage coil, and 32% in comparison with the local receive coil. Such SNR enhancement remains unchanged for different voxel sizes (larger and smaller FoVs). The analysis of metasurface effect on the image quality acquired with different acquisition matrices is included into Supplemental Material [42]). Hence, the proposed hybrid metasurface can be considered to be more efficient than the commercial coils currently used in clinical MRI. As was mentioned above, higher SNR potentially leads to more informative images or an ability to perform faster examinations with the same MRI machine.

We note that SNR decreases with the depth being a physical limitation of any non-volumetric coil. In the analysis above we considered one of the middle slices of the grapefruit (which was chosen from the multi-slice data set). The metasurface improves SNR in comparison with the local coil up to 8 cm in depth (see discussions and experimental results in Supplemental Material [42]).

## V. CONCLUSION AND FUTURE DIRECTIONS

We have proposed and studied experimentally a novel design of microwave metasurfaces capable of substantially enhancing image quality of commercial MRI machines. Due to a hybrid nature of the design, this metasurface provides a homogeneous enhancement of the RF magnetic field in the region-of-interest that is particularly relevant to MRI applications. Another distinct advantage of the suggested design is its ability to perform a fine tuning of the metasurface properties for patient-specific or object-specific investigations. It is worth noting that its eigenmode can be adjusted to a specific frequency of MRI operation even in the presence of different objects, and such a metasurface can be used for MRI of different nuclei. For the proof-of-principle experiment we changed the effective permittivity of the metasurface thus affecting the properties of the eigenmode, while in practice other advanced approaches to achieve such a tunability can be involved, e.g. photosensitive [44] and nonlinear meta-atoms [45]. The hybrid tunable metasurface acts as a wireless coil enhancing the SNR compared to a body coil transmit/receive more than 7 times in the region of interest leading to better image quality.

It is important to mention that there have been many alternative concepts proposed to increase the intrinsic sensitivity of MRI, such as, the use of multiple receivers [46], dielectric pads [47], passive resonant repeaters [48, 49], the dissolution dynamic nuclear polarization method [50], digitally controlled passive elements [51] and magneto-inductive waveguides [52]. The proposed metasurface-based approach can be seen very much as a complementary method that could be combined with the recently developed fully functional, printed and flexible MRI coils [53] or with structures based on very high permittivity materials [54], to provide an even greater practical impact. Equally well, the design of advanced electromagnetic structures based on meta-technologies can be regarded as the first-step towards completely novel devices with superior characteristics and techniques for clinical MRI applications.

**ACKNOWLEDGEMENTS**

This research was supported by Ministry of Education and Science of the Russian Federation (Zadanie No. 3.2465.2017/ПЧ) and ITMO University. The authors thank Dr. George Palikaras, Prof. Nico van den Berg and Dr. Alexander Raaijmakers for useful discussions, Prof. Vladimir Fokin, Dr. Alexander Efimtcev, and Andrey Sokolov for assistance with the MRI experiments. APS acknowledges a support by the IEEE MTT-S and Photonics Graduate Fellowships. AGW acknowledges support by European Research Council Advanced Grant 670629 NOMA MRI and NWO Topsubside.

**APPENDIX A: EXPERIMENTAL SETUP**

We have realized the hybrid metasurface as an array of 14x2 brass wires. Edges of the wires were inserted up to 25 mm into the plastic boxes, fillable by distilled water with permittivity $\varepsilon = 78+0.2i$ at a room temperature. First, we have performed measurements of the resonance frequency shift as a function of water level. For this purpose, the copper loop antenna was placed on top of the phantom (filled with tap water) located in the central part of the metasurface. The phantom parameters were: material $\varepsilon = 78+3.8i$, length = 130 mm, width = 90 mm and height = 80 mm. The structure was placed inside a 1.5 T Siemens Magnetom Espree MRI machine and we analysed the reflection coefficient of the loop antenna with the aid of a Planar TR1300/1 compact network analyser.

To obtain the dependence of the signal as a function of the tip angle (Fig. 2) the birdcage coil was used for transmitting and receiving the signal. We collected images for cases, with and without the metasurface, and calculated the SNR in each scenario. The SNR was determined by the average value of the signal in the region of the phantom divided by the standard deviation of the noise in the signal free area of the image. We used a spoiled gradient echo sequence with the following parameters: FoV 163x326 mm$^2$, TR/TE 1010/12.2 ms. Owing to the fact that the hybrid metasurface also enhances the RF transmitted magnetic field amplitude, each point in Figure 2a has been acquired for optimal RF power level, corresponding to the maximum of the signal, in order to avoid over-tipping (the phenomenon when $\lambda(r) \cdot \theta_{nominal} > \pi/2$) in the region of interest. By tuning the eigenmode frequency closer to the nuclear magnetic resonance frequency the RF power level was decreased. Also, it is important to note that in the case of the metasurface we have used distilled water as dielectric blocks, this slight influence on the level of the MR signal due to the water provides additional signal to the phantom signal. To properly account for this effect, for all scans without the metasurface we placed similar blocks with the water around the phantom. The data from Figure 3 were acquired in the same way.

To perform the image quality evaluation (see Fig. 4), we take a grapefruit that provides a rather good contrast in gradient-echo MR images. We consider one of the middle slices of the grapefruit (which is chosen from the multi-slice data set) at approximately 4 cm from the metasurface. A spoiled gradient-echo sequence with the following settings is used to perform MRI scans: acquisition matrix - 256x256; TR/TE 900/28 ms; optimal tip angle (corresponding to maximum of the signal) for the case with a metasurface - $14^0$, for the case without a metasurface - $75^0$, for local receive coil - $65^0$. For comparison with a commercial receive coil, we employ a 4-channel large flex receive coil from Siemens. The RF power level is adjusted in each case to obtain the maximum signal. To image different voxels sizes, we use FoV=180x180 mm$^2$, FoV=120x120 mm$^2$ and, in order to get the smaller voxel size, we reduce the FoV to 60x60 mm$^2$. Also it is important to notice that the signal strength decreases for images with smaller voxels size, so colorbars (Fig. 4) correspond only to the images with the same voxels size.

**APPENDIX B: NUMERICAL SIMULATIONS**

The numerical simulations are performed by using the Frequency Domain solver of CST Microwave Studio 2016. We use the plane wave excitation (propagating in *x*-direction with electric field polarized along the wires) and PML boundary conditions to calculate the eigenmode profile (Fig.1a). The used parameters of the metasurface are: an array 14x2, wire length L = 255 mm, period a = 10 mm and radius r = 1 mm. The wires are brass placed in a dielectric background material with $\varepsilon = 78$. The hybrid metasurface is placed in the background with $\varepsilon_1 = 1$ and the edges of wires (25 mm of metal from each side) are inserted into high-permittivity dielectric pads with $\varepsilon_2 = 78$ and the following geometrical dimensions (height (variable) = 106-198 mm; length = 130 mm; width = 380 mm). The metasurface is inserted in the centre of the dielectric pads. In order to shift the hybrid metasurface eigenmode closer to the 63.8 MHz, the wire length is increased to L = 331 mm.

In order to analyze the effect of metasurface tunability, we study the reflection coefficient of a loop antenna made from the brass material with the radius of 40 mm and wire diameter of 2 mm. To properly evaluate the resonance frequency shift, we include the MRI phantom to the simulations. The phantom is placed on top of the metasurface in the centre and is represented by the following parameters: material $\varepsilon = 78+0.2i$, length = 130 mm, width = 90 mm and height = 80 mm. The distance between the loop and surface of wires is 100 mm. In the numerical simulations, we also place our structure inside a copper cylinder with internal diameter of 589 mm and length of 1200 mm, which mimics the properties of a real scanner bore, to include the impact of a RF-shield of a scanner on the metasurface resonance frequency shift.